\documentclass[reprint]{revtex4-1}
\usepackage{graphicx} 
\usepackage{amsmath}
\usepackage{amssymb}
\usepackage{wasysym}

\begin{document}
\title{Probability of observing a number of unfolding events while stretching poly-proteins}

\author{Rodolfo I. Hermans}
\email{r.hermans@ucl.ac.uk}
\affiliation{London Centre for Nanotechnology, University College London}
\affiliation{Department of Physics and Astronomy, University College London}

\begin{abstract}
The mechanical stretching of single poly-proteins is an emerging tool for the study of protein (un)folding, chemical catalysis and polymer physics at the single molecule level.
The observed processes i.e unfolding or reduction events, are typically considered to be stochastic and by its nature are susceptible to be censored by the finite duration of the experiment.
Here we develop a formal analytical and experimental description on the number of observed events under various conditions of practical interest. We provide a rule of thumb to define the experiment protocol duration. Finally we provide a methodology to accurately estimate the number of stretched molecules based on the number of observed unfolding events. Using this analysis on experimental data we conclude for the first time that poly-ubiquitin binds at a random position both to the substrate and to the pulling probe and that observing all the existing modules is the less likely event.
\end{abstract}

\maketitle 

\section{Introduction}

Atomic force spectroscopy based single molecule force spectroscopy technique (AFM-SMFS) (described in detail elsewhere\cite{Fisher2000,Fernandez2004,Oberhauser2001,Baro2012}) allows calibrated mechanical stretching and monitoring of individual polymer molecules such as sugars, DNA or proteins. For a decade this tool has been used for the study of the folding and unfolding mechanism of proteins at the single molecule level and lately it has emerged as the means to test chemical catalysis at the level of a localized single disulfide bond~\cite{Perez-Jimenez2011592,Garcia-Manyes2011}.
In order to obtain a suitable `fingerprint' \emph{i.e.} unambiguous independent evidence of the controlled condition of the intended sample, protein samples are often engineered to contain multiple well separated identical repeats of the structure of interest~\cite{CarrionVazquez2000}. The mechanics of these poly-proteins is expected to be identical to monomers but data from poly-proteins is considered to be less likely to be perturbed by spurious surface interactions~\cite{Garcia-Manyes2007a}.
Its has been observed experimentally that for a poly-protein designed to contain $N$ identical modules, the number of observed modules is prevailingly smaller than $N$, but to our knowledge there hasn't been a thorough analytical explanation for this phenomena. The following analysis explain the experimentally observed distribution of events by considering the stochastic nature of three relevant processes that define the experiment: the bonding of the sample, the unfolding event and the spontaneous (or programmed) termination of the experiment. The analysis is valid in a broad scope and the experimental data is provided as an representative example.

\section{Experimental observation}

     \begin{figure}[tbh]
        \begin{center}
        \includegraphics[width=\columnwidth]{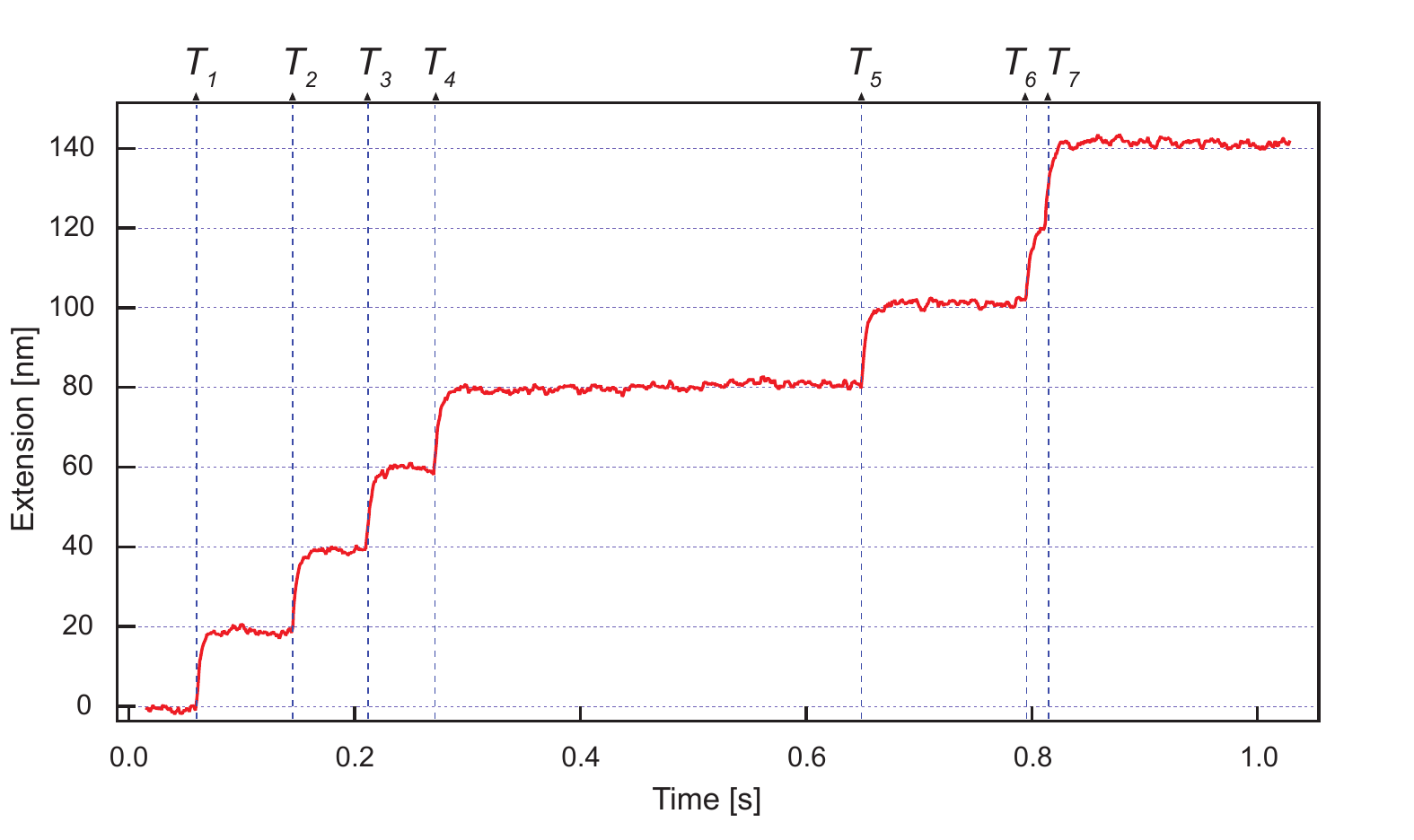}\\
        \end{center}
        \caption{Example of the raw data in an AFM-SMFS experiment. End-to-end length of the protein sample features extension events at times $\{T_1, T_2, \ldots, T_7\}$, each one corresponding to the unfolding of a protein module. The experiment is interrupted shortly after 1 second, so any further events are `censored'.}
      \end{figure}

 We study single poly-protein chains engineered to contain exactly $N_{\text{max}}=12$ identical domains of the protein ubiquitin (Ubi12), stretched by AFM at a constant force of $110$~pN~\cite{Brujic2006a}. The stretching force weakens the tertiary structure of all ``protein modules" exposed to force allowing them to unfold, yet the number of such observed unfolding events varies from case to case and rarely reached the total of 12 engineered modules.  Figure~1 shows an example of the raw data obtained by AFM-SMFS in constant-force mode, featuring only $k_{\text{max}}=7$ unfolding events at times $\{T_1, T_2, \ldots, T_7\}$, in an experiment that lasted around $1$ second. \mbox{Figure 2} shows the population distribution  of the number of observed unfolding events in each single poly-protein chain for an unfiltered data set consisting of $1198$ unfolding poly-protein chains and a total of $1741$ unfolding events. The number of traces observed to unfold $k_{\text{max}}$ events seems to decrease approximately by $22\%$ for each increment in $k_{\text{max}}$.

     \begin{figure}[bt]
        \begin{center}
        \includegraphics[width=\columnwidth]{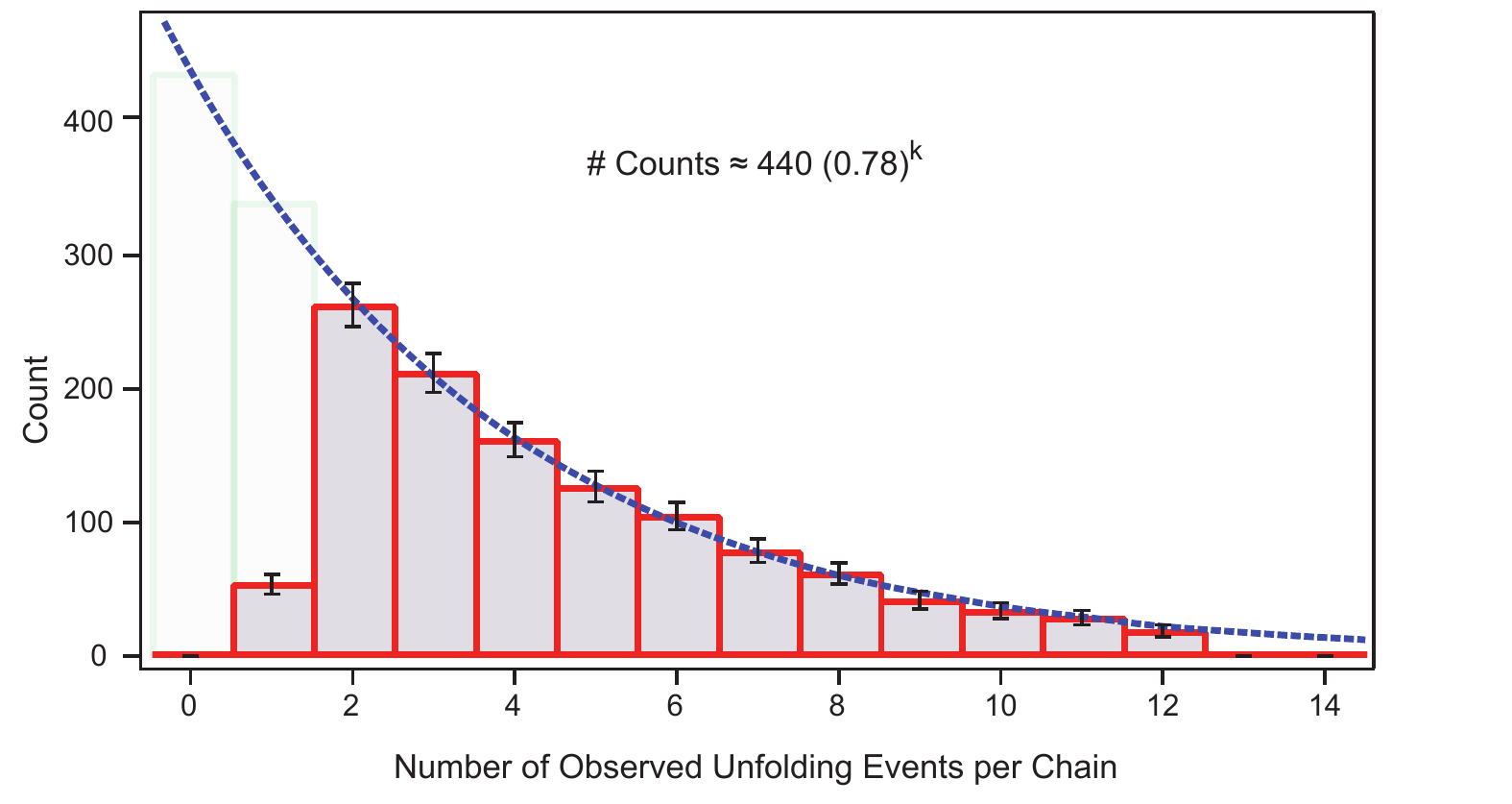}\\
        \end{center}
        \caption{Number of observed unfolding events in each single poly-protein chain. Despite the fact that the poly-ubiquitin chain is designed to contain 12 identical modules, the population of observed long chains is much smaller than short ones. Except for traces with zero or one unfolding events, the trend of observed number of events versus chain length can be empirically modeled with the function $(0.78)^{k}$, implying that the population of observed events decreases by $22\%$ for each extra observed unfolding module.}
      \end{figure}

Here we explain this observed distribution with a simple approach that does not require assuming a particular model for the unfolding kinetics and relies only on the assumptions that the unfolding events are independent and identically distributed.

\section{Probability of observing the $k^{\text{th}}$ event out of $N$ }

  We assume a chain of $N$ identical modules and define $t=0$ when the stretching force is first applied. The unfolding events occur between time $0$ and $t$ with probability $P_e(t)$. The experiment is interrupted stochastically by spontaneous detachment of the sample with probability density $p_i(t)$ or by design at a known time. The experimental measurables are the observed unfolding times $T_k$ with $0\leq T_1 \leq T_2 \leq,\ldots,T_{k_{\text{max}}}$, and the censoring time $T_i$ that is the event of the experiment been interrupted at time $T_i>T_{k_{\text{max}}}$. Any possible subsequent event after $T_i$ is not measurable, constituting a \emph{type 2 right censored} system\cite{HandbookStatisticsV16}.

  Now we obtain the analytical expression of the probability for a $k^{\text{th}}$ order statistics\cite{David2006} to be the last one observed before time $t$. The probability of a series of $k$  modules to unfold before the time $t$ is $\left(P_e(t)\right)^k$. This $k$ modules can be chosen in $\binom{N}{k}$ different ways. For the $k^{\text{th}}$ to be the last event, no other events should happen after the interruption of the experiment. Then the probability that \emph{any} $k$ modules unfolded before $t$ and the remaining $N-k$ did not, is given by:

   \begin{equation}
   P(T_{k_{\text{max}}}<t \, |N)\,  = \,\frac{N!}{k!\,(N-k)!}\,\, P_e(t)^k \,\,(1-P_e(t))^{N-k} \, .\nonumber
  \end{equation}

   Now we consider that the experiment is interrupted between time $t$ and $t+dt$ with probability $p_i(t) dt$, now the probability that the $k^{\text{th}}$ is the last observed event at any time is

\begin{eqnarray}   \label{eq:PKmaxGN}
  p(k|N) &=&  \int_0^{\infty} p_i(t) \,  P(T_{k_{\text{max}}}<t \, |N)\,dt \nonumber \\
   &=& \frac{N!}{k!\,(N-k)!} \, \int_0^{\infty} p_i(t) \,  P_e(t)^k \,(1-P_e(t))^{N-k}\,dt  \, .\nonumber \\
\end{eqnarray}

 If algebraic expressions for $p_i(t)$ and $P_e(t)$ are available, equation~1 could be integrated analytically. For example, a trivial case of practical interest is the idealized Arrhenius kinetics when the unfolding rate is a single exponential $P_e(t)= 1-e^{-\alpha t}$ and experiment is only interrupted by design at time $t_d$, and $p_i(t) = \delta (t-t_d)$. By making $k=N$ we see that the probability of observing the last event out of $N$ is:

 \begin{equation}\label{eq:ideal}
    p_{\text{A}}(N|N)= {\left(1-e^{-\alpha  t_d}\right)}^N\, .
 \end{equation}

 Solving for $t_d$ allows the design of an experimental protocol that extends in time enough to observe a desired proportion of the last event~\cite{Hermans2010}.

 Another example of interest is the case where the experiment is interrupted by the spontaneous detachment of the sample from the substrate or the cantilever tip. If we assume $p_i(t)= -\beta e^{-\beta t}$  in equation~1 we obtain

  \begin{equation}\label{eq:Gamma}
 p(k|N)= \frac{\beta}{\alpha} \frac{n!}{(n-k)!} \frac{\Gamma \left(-k+n+\frac{\beta }{\alpha }\right)}{\Gamma \left(n+\frac{\beta }{\alpha }+1\right)}
  \end{equation}

  where $\Gamma(z)$ is Euler's Gamma function.  We choose to avoid any model assumptions and integrate equation~\ref{eq:PKmaxGN} numerically by estimating $p_i(t)$ and $P_e(t)$ directly from the raw experimental data.

 \section{Data analysis}

 We assume the experimental data has been analyzed and compiled in $R$ sets containing the unfolding event times $\{T_1, T_2, \ldots, T_{k_{\text{max}}}\}$ and the censoring times $T_d$.
 All the events times and censoring times are respectively aggregated to estimate the their probability functions. The cumulative distribution function (CDF) is calculated directly by an interpolating formula.

     \begin{equation}\label{eq:CPFinterp}
       \hat{P}(t) = C\left(k(t)+\frac{t-t_k}{t_{k+1}-t_k}\right) \, ,
     \end{equation}

 where  $k(t)$ is the rank of the sample $t_k$, such that $t\in [t_k,t_{k+1}]$, and $C$ is such that $\displaystyle{\lim_{t\to\infty} \hat{P}(t) = 1}$. The probability density function (PDF) is calculated by a Gaussian kernel density estimation~\cite{Gerard1999}.  Both CDF and PDF calculations are already implemented in a simple to use function \emph{SmoothKernelDistribution} in Wolfram's Mathematica~\cite{Research2013}.
 
      \begin{table}[tb]
        \begin{center}
        \includegraphics[width=\columnwidth]{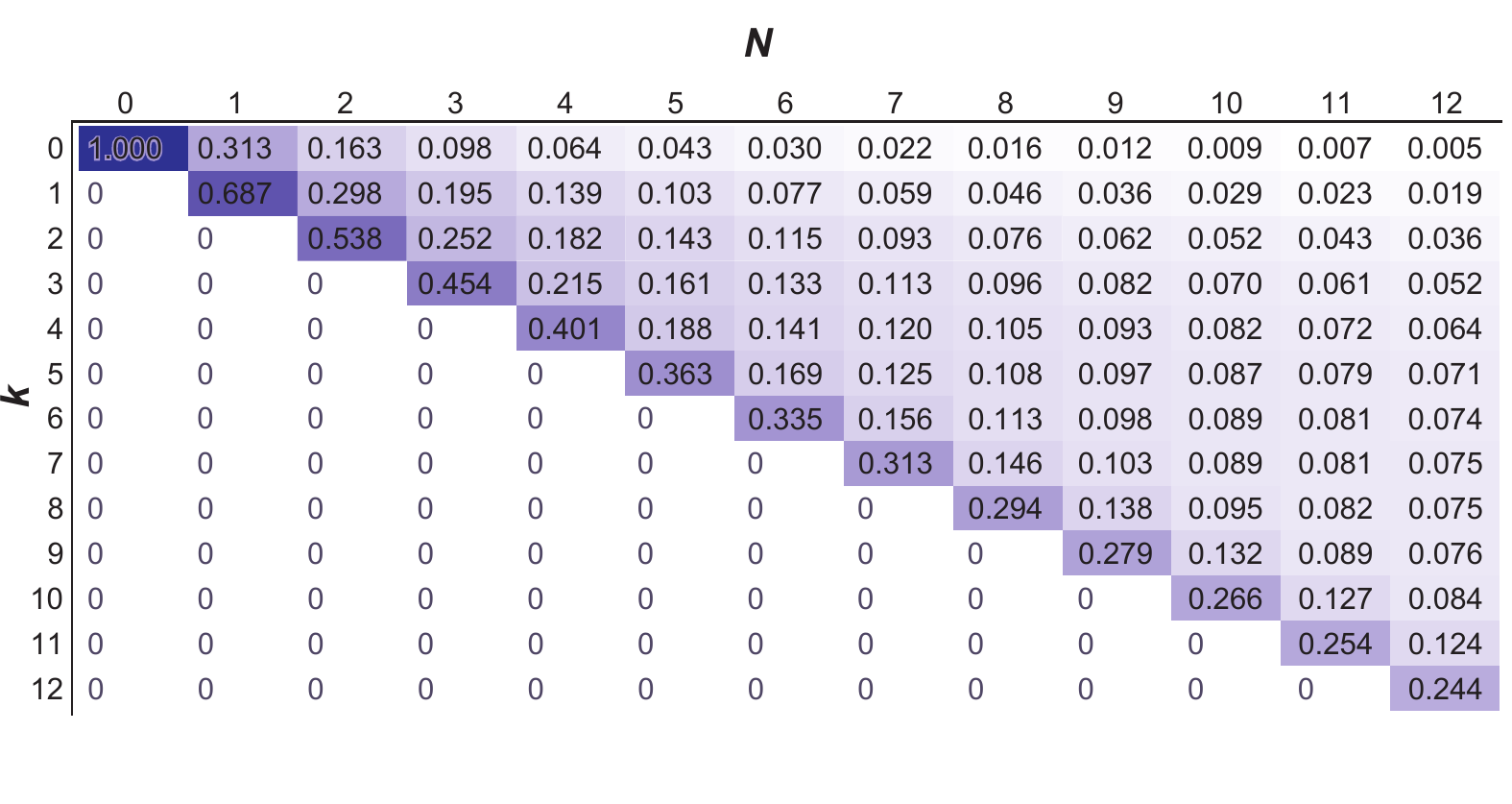}\\
        \end{center}
        \caption{Matrix for the probability of observing $k$ out of $N$ events obtained from the experimental probabilities of unfolding and detachment.}
      \end{table}

The probability $p(k|N)$ is calculated numerically by integrating equation~\ref{eq:PKmaxGN} using the estimated $P_e(t)$ from equation~\ref{eq:CPFinterp}. The calculated values are given in table~1.  The values of $p(k_{\text{max}}|N)$ in table 1 are arranged in a matrix $\mathbf{M}_{k,n}$ for algebraic convenience.   A consequence of censoring is that observing $k_{\text{max}}$ events is possible from from traces with any value of $N$ in the range $N_{\text{max}} \geq N \geq k_{\text{max}}$. Consequently the overall probability of observing $k_{\text{max}}$ events  $P(k_{\text{max}})$ is given by

  \begin{equation}\label{eq:PkMax}
    P(k_{\text{max}}) = \sum_{N=1}^{N_{\text{max}}} \, p(k_{\text{max}}|N)\, P(N) \, .
  \end{equation}

  We new define the vectors $\vec{P}_k$ and $\vec{P}_n$ such that their components are the probabilities $P(k_{\text{max}})$ and  $P(N)$ respectively.

\begin{eqnarray}
  \vec{P}_k &=& \{\,P(k_{\text{max}}=1), \, P(k_{\text{max}}=2), \, \ldots, \, P(k_{\text{max}}=N_{\text{max}}) \, \} \nonumber \\
  \vec{P}_n &=& \{\,P(N=1), \, P(N=2), \, \ldots, \, P(N=N_{\text{max}}) \, \} \nonumber
\end{eqnarray}

   Using this definition equation~\ref{eq:PkMax} can be written more conveniently in matrix form

 \begin{equation}\label{eq:TransformPopulations}
   \vec{P}_k = \,\mathbf{M}_{k,n}\, \vec{P}_n \, .
 \end{equation}

    Matrix $\mathbf{M}_{k,n}$ allows us to transform the probability (or population) distributions from the available modules $N$ to the observed modules $k_{\text{max}}$, and \emph{vice versa}. Now we investigate the expected population of observed number of events per chain for three different significant scenarios  represented in figures~3a, 3c and 3e.

\begin{enumerate}
  \item All poly-protein chain are deterministically picked from the end linkers, exposing all 12 modules to force. $P(N) = \delta(12)$ (Figure~3a)
  \item One of the end linkers is fixed to the substrate and the cantilever picks randomly the chain at any linker, exposing between one and $12$ modules with equal probability. $P(12\geq N >0) = 1/12$ (Figure~3c)
  \item Both ends are picked randomly. There are $(13-N)$ ways to pick $N$ modules, making a total of $12\times 11/2 = 66$. $P(12\geq N >0) = (13-N)/66$ (Figure~3e)
\end{enumerate}

      \begin{figure}[tb]
        \begin{center}
        \includegraphics[width=0.99\columnwidth]{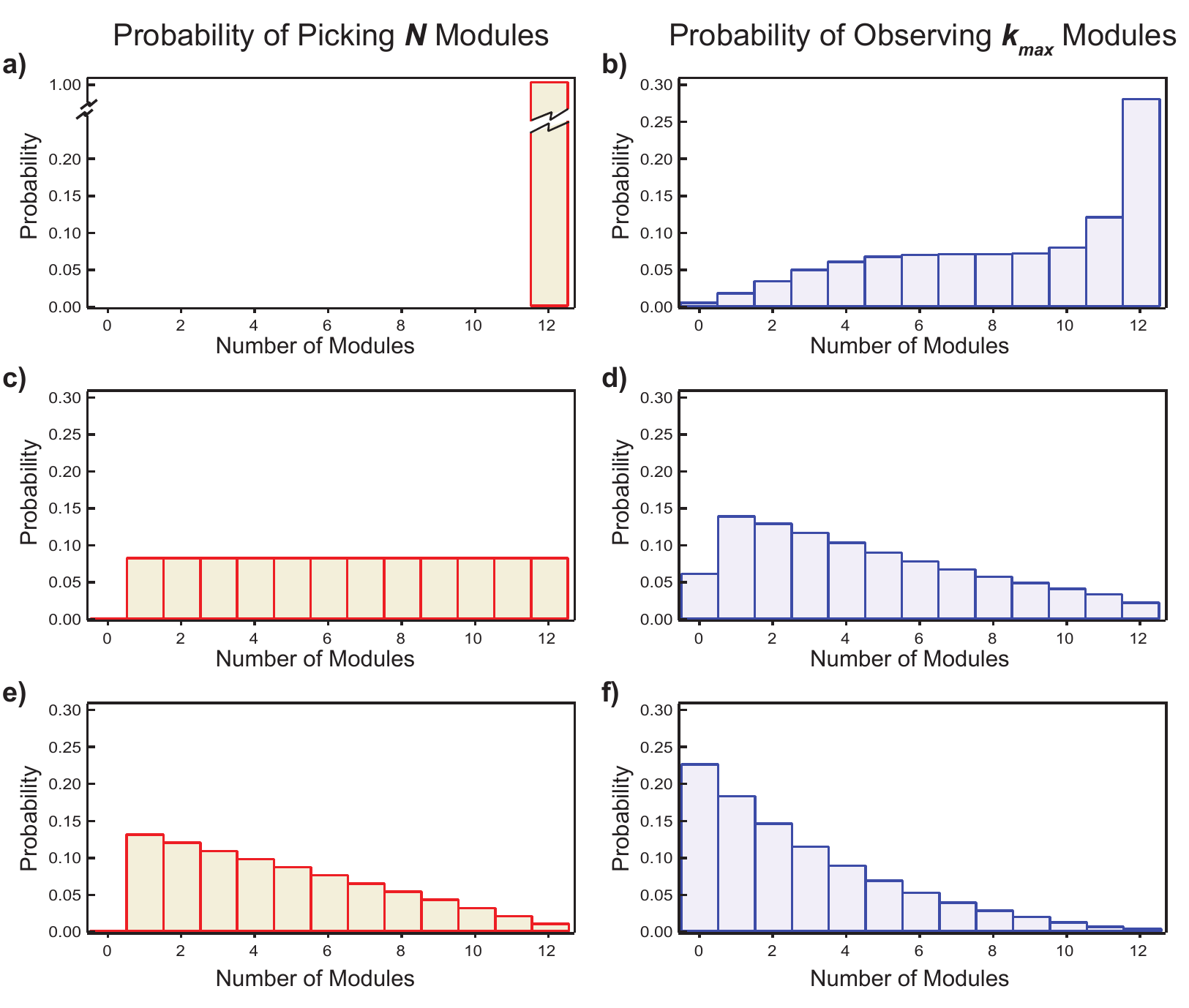}\\
        \end{center}
        \caption{Probability of picking $N$ (left) or observing $k$ (right) protein modules under three different scenarios. Protein grabbed from a,b) the ends (always), c,d) one end fixed and one  random position, e,f) two random places in the chain }
      \end{figure}

The calculated probability of observing $k_{\text{max}}$ modules is shown in figures~3b, 3d and 3e. We can notice that figure~3e looks remarkably similar to the histogram of experimentally observed $k_{\text{max}}$ in figure~2, except for the first two bins corresponding to $k_{\text{max}}=0$ and $k_{\text{max}}=1$.

The small number of counts in the first two bins of  figure~2 is explained by the fact that traces with less than two events are rarely saved because of the lack of a characteristic fingerprint to differentiate them from non specific sample.

      \begin{figure}[tb]
        \begin{center}
        \includegraphics[width=0.99\columnwidth]{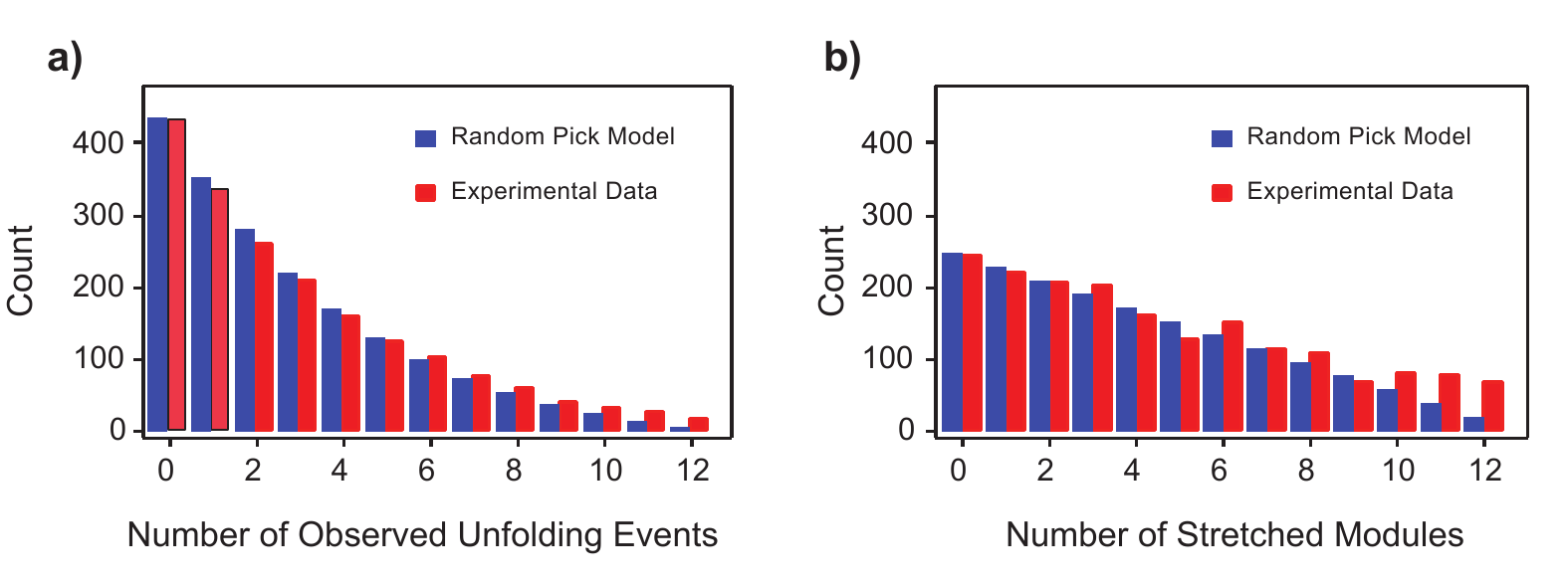}\\
        \end{center}
        \caption{Comparison between observed distribution of chain length with model of random attachment places}
      \end{figure}

Now we calculate  $\mathbf{M}_{k,n}^{-1}$, the inverse of $\mathbf{M}_{k,n}$ to solve the inverse problem and estimate the population distribution of $N$ based on the population of $k_{\text{max}}$.

 \begin{equation}
   \vec{P}_n = \mathbf{M}_{k,n}^{-1} \, \vec{P}_k
 \end{equation}

In order to  compensate for the censored data in the first two bins of  figure~2, we extrapolate the count values for $k_{\text{max}}=0$ and $k_{\text{max}}=1$ to match the observed trend using an exponential fit\footnote{Notice that if the two fist bins in figure~2 were not corrected then the estimated population of $N$ would contain negative values.}. The corrected histogram is plotted with red bars in figure~4a. For comparison in the same figure the blue bars indicate the distribution expected when both ends of the poly-protein are picked randomly (Scenario 3).

The calculated population of $N$ is shown in figure~4b in red bars, together with the distribution expected when both ends of the poly-protein are picked randomly in blue bars. Figures~4a and~4b confirms the similarity of the experimental data and the expected distribution for random binding of the poly-protein.

From this result, we can conclude that proteins are anchored in a completely random process, both to the substrate and to the cantilever. Even if any anchor position is equally probable, the fact that there is only one way to pick all $12$ modules but several more more ways to pick a smaller number of modules explains the experiential distribution $\vec{P}_n$ of the number $N$ of stretched modules. The population of observed events $\vec{P}_k$ is further reduced by censoring.

\section{Discussion}

The procedure shown is a quantification of the probability of censoring data and allows to transform the observed population of events into the existing population based on a matrix constructed numerically based on experimental data and free of any assumptions on the kinetics of the events. The calculated existing population is consistent with the scenario where both anchoring points are chosen among all the protein modules with equal probability and the probability of picking $N$ modules is $P(N)=2 (N_{\text{max}}-N+1)/(N_{\text{max}}(N_{\text{max}}-1))$. Therefore, the less likely configuration is to have a protein grabbed from the ends, and the most likely is to grab a single module. In presence of spontaneous detachment of the sample or any other limitation on the experiment duration, the observed distribution is further skewed by censoring last events making the observation of the unfolding of all engineered modules relatively unlikely.

In the presented experimental the protein sample was attached non-specifically to the substrate and AFM cantilever tip and we concluded that the stretching ends were picked randomly. That scenario may not be the one present on experiments performed with specific functionalization at the protein ends. Specific treatment of substrate and cantilever surfaces may be desirable to allow grabbing polymer samples form the ends or at least reduce spontaneous detachment of the sample and consequently minimize censoring.

The importance of addressing censoring in order to obtain unbiased estimators has been previously addressed for different single-molecules experiments~\cite{Koster07022006}. The use of non-parametric estimators such as maximum likelihood~\cite{Brujic2006a} that addresses all the de details of the detection process  is of paramount importance. Ignoring that a finite experiment duration inevitably leads to censoring has lead Cao \emph{et.al.} to mistakenly conclude that, under experimental conditions, the observed unfolding dwell times does not depend on the number $N$ of available modules~\cite{Cao2011}.

%

\section{Conclusions}
We have identified the study of poly-proteins by AFM-SMFS as a type-2 right censored system and provided a model-independent method to understand and predict the population of observed unfolding events. The population of modules exposed to force is accurately estimated based in the number of observed events accounting for the unavoidable censoring of the data. Using this method we concluded that the experimental data on AFM-SMFS unfolding of Ubi12 stretched at constant $110$ pN clearly suggest that the attachment of the protein sample to the substrate and pulling probe is random and therefor observing the unfolding of all possible modules is highly unlikely. We are confident that the tools here developed will also be of use to the experimentalist in the process of designing protocols that account for the often ignored stochastic nature of the attachment and detachment of the sample to substrate and probe.

\begin{acknowledgments}
This work is inspired by previous work during my Ph.D. studies. Thanks to my former Ph.D. advisor Professor Julio M. Fernandez for access to the experimental data.
I thank also Dr. Sergi Garcia-Manyes and Mr. Samir Aoudjane for valuable discussion.
\end{acknowledgments}


\begin{thebibliography}{17}%
\makeatletter
\providecommand \@ifxundefined [1]{%
 \@ifx{#1\undefined}
}%
\providecommand \@ifnum [1]{%
 \ifnum #1\expandafter \@firstoftwo
 \else \expandafter \@secondoftwo
 \fi
}%
\providecommand \@ifx [1]{%
 \ifx #1\expandafter \@firstoftwo
 \else \expandafter \@secondoftwo
 \fi
}%
\providecommand \natexlab [1]{#1}%
\providecommand \enquote  [1]{``#1''}%
\providecommand \bibnamefont  [1]{#1}%
\providecommand \bibfnamefont [1]{#1}%
\providecommand \citenamefont [1]{#1}%
\providecommand \href@noop [0]{\@secondoftwo}%
\providecommand \href [0]{\begingroup \@sanitize@url \@href}%
\providecommand \@href[1]{\@@startlink{#1}\@@href}%
\providecommand \@@href[1]{\endgroup#1\@@endlink}%
\providecommand \@sanitize@url [0]{\catcode `\\12\catcode `\$12\catcode
  `\&12\catcode `\#12\catcode `\^12\catcode `\_12\catcode `\%12\relax}%
\providecommand \@@startlink[1]{}%
\providecommand \@@endlink[0]{}%
\providecommand \url  [0]{\begingroup\@sanitize@url \@url }%
\providecommand \@url [1]{\endgroup\@href {#1}{\urlprefix }}%
\providecommand \urlprefix  [0]{URL }%
\providecommand \Eprint [0]{\href }%
\providecommand \doibase [0]{http://dx.doi.org/}%
\providecommand \selectlanguage [0]{\@gobble}%
\providecommand \bibinfo  [0]{\@secondoftwo}%
\providecommand \bibfield  [0]{\@secondoftwo}%
\providecommand \translation [1]{[#1]}%
\providecommand \BibitemOpen [0]{}%
\providecommand \bibitemStop [0]{}%
\providecommand \bibitemNoStop [0]{.\EOS\space}%
\providecommand \EOS [0]{\spacefactor3000\relax}%
\providecommand \BibitemShut  [1]{\csname bibitem#1\endcsname}%
\let\auto@bib@innerbib\@empty
\bibitem [{\citenamefont {Fisher}\ \emph {et~al.}(2000)\citenamefont {Fisher},
  \citenamefont {Marszalek},\ and\ \citenamefont {Fernandez}}]{Fisher2000}%
  \BibitemOpen
  \bibfield  {author} {\bibinfo {author} {\bibfnamefont {T.~E.}\ \bibnamefont
  {Fisher}}, \bibinfo {author} {\bibfnamefont {P.~E.}\ \bibnamefont
  {Marszalek}}, \ and\ \bibinfo {author} {\bibfnamefont {J.~M.}\ \bibnamefont
  {Fernandez}},\ }\href {http://dx.doi.org/10.1038/78936} {\bibfield  {journal}
  {\bibinfo  {journal} {Nat Struct Mol Biol}\ }\textbf {\bibinfo {volume}
  {7}},\ \bibinfo {pages} {719} (\bibinfo {year} {2000})}\BibitemShut {NoStop}%
\bibitem [{\citenamefont {Fernandez}\ and\ \citenamefont
  {Li}(2004)}]{Fernandez2004}%
  \BibitemOpen
  \bibfield  {author} {\bibinfo {author} {\bibfnamefont {J.~M.}\ \bibnamefont
  {Fernandez}}\ and\ \bibinfo {author} {\bibfnamefont {H.}~\bibnamefont {Li}},\
  }\href {\doibase 10.1126/science.1092497} {\bibfield  {journal} {\bibinfo
  {journal} {Science}\ }\textbf {\bibinfo {volume} {303}},\ \bibinfo {pages}
  {1674} (\bibinfo {year} {2004})}\BibitemShut {NoStop}%
\bibitem [{\citenamefont {Oberhauser}\ \emph {et~al.}(2001)\citenamefont
  {Oberhauser}, \citenamefont {Hansma}, \citenamefont {Carrion-Vazquez},\ and\
  \citenamefont {Fernandez}}]{Oberhauser2001}%
  \BibitemOpen
  \bibfield  {author} {\bibinfo {author} {\bibfnamefont {A.~F.}\ \bibnamefont
  {Oberhauser}}, \bibinfo {author} {\bibfnamefont {P.~K.}\ \bibnamefont
  {Hansma}}, \bibinfo {author} {\bibfnamefont {M.}~\bibnamefont
  {Carrion-Vazquez}}, \ and\ \bibinfo {author} {\bibfnamefont {J.~M.}\
  \bibnamefont {Fernandez}},\ }\href
  {http://www.pnas.org/content/98/2/468.abstract} {\bibfield  {journal}
  {\bibinfo  {journal} {Proceedings of the National Academy of Sciences of the
  United States of America}\ }\textbf {\bibinfo {volume} {98}},\ \bibinfo
  {pages} {468} (\bibinfo {year} {2001})}\BibitemShut {NoStop}%
\bibitem [{\citenamefont {Bar\'{o}}\ and\ \citenamefont
  {Reifenberger}(2012)}]{Baro2012}%
  \BibitemOpen
  \bibinfo {editor} {\bibfnamefont {A.~M.}\ \bibnamefont {Bar\'{o}}}\ and\
  \bibinfo {editor} {\bibfnamefont {R.~G.}\ \bibnamefont {Reifenberger}},\
  eds.,\ \href
  {http://eu.wiley.com/WileyCDA/WileyTitle/productCd-3527327584,subjectCd-PHD0.html}
  {\emph {\bibinfo {title} {{Atomic Force Microscopy in Liquid: Biological
  Applications}}}}\ (\bibinfo  {publisher} {John Wiley \& Sons},\ \bibinfo
  {year} {2012})\ p.\ \bibinfo {pages} {402}\BibitemShut {NoStop}%
\bibitem [{\citenamefont {Perez-Jimenez}\ \emph {et~al.}(2011)\citenamefont
  {Perez-Jimenez}, \citenamefont {Ingl\'{e}s-Prieto}, \citenamefont {Zhao},
  \citenamefont {Sanchez-Romero}, \citenamefont {Alegre-Cebollada},
  \citenamefont {Kosuri}, \citenamefont {Garcia-Manyes}, \citenamefont
  {Kappock}, \citenamefont {Tanokura}, \citenamefont {Holmgren}, \citenamefont
  {Sanchez-Ruiz}, \citenamefont {Gaucher},\ and\ \citenamefont
  {Fern\'{a}ndez}}]{Perez-Jimenez2011592}%
  \BibitemOpen
  \bibfield  {author} {\bibinfo {author} {\bibfnamefont {R.}~\bibnamefont
  {Perez-Jimenez}}, \bibinfo {author} {\bibfnamefont {A.}~\bibnamefont
  {Ingl\'{e}s-Prieto}}, \bibinfo {author} {\bibfnamefont {Z.-M.}\ \bibnamefont
  {Zhao}}, \bibinfo {author} {\bibfnamefont {I.}~\bibnamefont
  {Sanchez-Romero}}, \bibinfo {author} {\bibfnamefont {J.}~\bibnamefont
  {Alegre-Cebollada}}, \bibinfo {author} {\bibfnamefont {P.}~\bibnamefont
  {Kosuri}}, \bibinfo {author} {\bibfnamefont {S.}~\bibnamefont
  {Garcia-Manyes}}, \bibinfo {author} {\bibfnamefont {T.~J.}\ \bibnamefont
  {Kappock}}, \bibinfo {author} {\bibfnamefont {M.}~\bibnamefont {Tanokura}},
  \bibinfo {author} {\bibfnamefont {A.}~\bibnamefont {Holmgren}}, \bibinfo
  {author} {\bibfnamefont {J.~M.}\ \bibnamefont {Sanchez-Ruiz}}, \bibinfo
  {author} {\bibfnamefont {E.~A.}\ \bibnamefont {Gaucher}}, \ and\ \bibinfo
  {author} {\bibfnamefont {J.~M.}\ \bibnamefont {Fern\'{a}ndez}},\ }\href
  {http://www.scopus.com/inward/record.url?eid=2-s2.0-79955640296\&partnerID=40\&md5=dfdcb00eccfd305b810320812deab357}
  {\bibfield  {journal} {\bibinfo  {journal} {Nature Structural and Molecular
  Biology}\ }\textbf {\bibinfo {volume} {18}},\ \bibinfo {pages} {592}
  (\bibinfo {year} {2011})}\BibitemShut {NoStop}%
\bibitem [{\citenamefont {{Garcia-Manyes S.}}(2011)}]{Garcia-Manyes2011}%
  \BibitemOpen
  \bibfield  {author} {\bibinfo {author} {\bibfnamefont {K.~T.-L. F. J.~M.}\
  \bibnamefont {{Garcia-Manyes S.}}},\ }\href
  {http://www.scopus.com/inward/record.url?eid=2-s2.0-79952273128\&partnerID=40\&md5=80da2a5351e719b8849fa03b170a2932}
  {\bibfield  {journal} {\bibinfo  {journal} {Journal of the American Chemical
  Society}\ }\textbf {\bibinfo {volume} {133}},\ \bibinfo {pages} {3104}
  (\bibinfo {year} {2011})}\BibitemShut {NoStop}%
\bibitem [{\citenamefont {Carrion-Vazquez}\ \emph {et~al.}(2000)\citenamefont
  {Carrion-Vazquez}, \citenamefont {Oberhauser}, \citenamefont {Fisher},
  \citenamefont {Marszalek}, \citenamefont {Li},\ and\ \citenamefont
  {Fernandez}}]{CarrionVazquez2000}%
  \BibitemOpen
  \bibfield  {author} {\bibinfo {author} {\bibfnamefont {M.}~\bibnamefont
  {Carrion-Vazquez}}, \bibinfo {author} {\bibfnamefont {A.~F.}\ \bibnamefont
  {Oberhauser}}, \bibinfo {author} {\bibfnamefont {T.~E.}\ \bibnamefont
  {Fisher}}, \bibinfo {author} {\bibfnamefont {P.~E.}\ \bibnamefont
  {Marszalek}}, \bibinfo {author} {\bibfnamefont {H.}~\bibnamefont {Li}}, \
  and\ \bibinfo {author} {\bibfnamefont {J.~M.}\ \bibnamefont {Fernandez}},\
  }\href {\doibase DOI: 10.1016/S0079-6107(00)00017-1} {\bibfield  {journal}
  {\bibinfo  {journal} {Progress in Biophysics and Molecular Biology}\ }\textbf
  {\bibinfo {volume} {74}},\ \bibinfo {pages} {63} (\bibinfo {year}
  {2000})}\BibitemShut {NoStop}%
\bibitem [{\citenamefont {Garcia-Manyes}\ \emph {et~al.}(2007)\citenamefont
  {Garcia-Manyes}, \citenamefont {Brujic}, \citenamefont {Badilla},\ and\
  \citenamefont {Fern\'{a}ndez}}]{Garcia-Manyes2007a}%
  \BibitemOpen
  \bibfield  {author} {\bibinfo {author} {\bibfnamefont {S.}~\bibnamefont
  {Garcia-Manyes}}, \bibinfo {author} {\bibfnamefont {J.}~\bibnamefont
  {Brujic}}, \bibinfo {author} {\bibfnamefont {C.~L.}\ \bibnamefont {Badilla}},
  \ and\ \bibinfo {author} {\bibfnamefont {J.~M.}\ \bibnamefont
  {Fern\'{a}ndez}},\ }\href {\doibase DOI: 10.1529/biophysj.107.104422}
  {\bibfield  {journal} {\bibinfo  {journal} {Biophysical Journal}\ }\textbf
  {\bibinfo {volume} {93}},\ \bibinfo {pages} {2436} (\bibinfo {year}
  {2007})}\BibitemShut {NoStop}%
\bibitem [{\citenamefont {Brujic}\ \emph {et~al.}(2006)\citenamefont {Brujic},
  \citenamefont {{Hermans Z.}}, \citenamefont {Walther}, \citenamefont
  {Fernandez}, \citenamefont {{Hermans Z}},\ and\ \citenamefont
  {Bruji\'{c}}}]{Brujic2006a}%
  \BibitemOpen
  \bibfield  {author} {\bibinfo {author} {\bibfnamefont {J.}~\bibnamefont
  {Brujic}}, \bibinfo {author} {\bibfnamefont {R.~I.}\ \bibnamefont {{Hermans
  Z.}}}, \bibinfo {author} {\bibfnamefont {K.~a.}\ \bibnamefont {Walther}},
  \bibinfo {author} {\bibfnamefont {J.~M.}\ \bibnamefont {Fernandez}}, \bibinfo
  {author} {\bibfnamefont {R.~I.}\ \bibnamefont {{Hermans Z}}}, \ and\ \bibinfo
  {author} {\bibfnamefont {J.}~\bibnamefont {Bruji\'{c}}},\ }\href {\doibase
  10.1038/nphys269} {\bibfield  {journal} {\bibinfo  {journal} {Nature
  Physics}\ }\textbf {\bibinfo {volume} {2}},\ \bibinfo {pages} {282} (\bibinfo
  {year} {2006})}\BibitemShut {NoStop}%
\bibitem [{\citenamefont {Balakrishnan}(1998)}]{HandbookStatisticsV16}%
  \BibitemOpen
  \bibfield  {author} {\bibinfo {author} {\bibfnamefont {N.}~\bibnamefont
  {Balakrishnan}},\ }\href
  {http://www.elsevier.com/wps/find/bookdescription.cws\_home/600936/description\#description}
  {\emph {\bibinfo {title} {{Handbook of statistics, 16, Order Statistics:
  Theory \& Methods}}}}\ (\bibinfo  {publisher} {Elsevier Science Pub Co},\
  \bibinfo {year} {1998})\BibitemShut {NoStop}%
\bibitem [{\citenamefont {David}(2006)}]{David2006}%
  \BibitemOpen
  \bibfield  {author} {\bibinfo {author} {\bibfnamefont {H.}~\bibnamefont
  {David}},\ }\href {\doibase 10.1007/0-8176-4487-3\_10} {\emph {\bibinfo
  {title} {Advances in Distribution Theory, Order Statistics, and
  Inference}}},\ \bibinfo {series} {Statistics for Industry and Technology},
  Vol.\ \bibinfo {volume} {Topics in}\ (\bibinfo  {publisher} {Birkh\"{a}user
  Boston},\ \bibinfo {year} {2006})\ pp.\ \bibinfo {pages}
  {157--172}\BibitemShut {NoStop}%
\bibitem [{\citenamefont {Hermans}(2010)}]{Hermans2010}%
  \BibitemOpen
  \bibfield  {author} {\bibinfo {author} {\bibfnamefont {R.~I.}\ \bibnamefont
  {Hermans}},\ }\emph {\bibinfo {title} {{Experimental study of single protein
  mechanics and protein rates of unfolding}}},\ \href@noop {} {Ph.D. thesis},\
  \bibinfo  {school} {Graduate School of Arts and Sciences, Columbia
  University} (\bibinfo {year} {2010})\BibitemShut {NoStop}%
\bibitem [{\citenamefont {Gerard}\ and\ \citenamefont
  {Schucany}(1999)}]{Gerard1999}%
  \BibitemOpen
  \bibfield  {author} {\bibinfo {author} {\bibfnamefont {P.~D.}\ \bibnamefont
  {Gerard}}\ and\ \bibinfo {author} {\bibfnamefont {W.~R.}\ \bibnamefont
  {Schucany}},\ }\href {\doibase 10.1111/j.0006-341X.1999.00769.x} {\bibfield
  {journal} {\bibinfo  {journal} {Biometrics}\ }\textbf {\bibinfo {volume}
  {55}},\ \bibinfo {pages} {769} (\bibinfo {year} {1999})}\BibitemShut
  {NoStop}%
\bibitem [{\citenamefont {Research}(2013)}]{Research2013}%
  \BibitemOpen
  \bibfield  {author} {\bibinfo {author} {\bibfnamefont {W.}~\bibnamefont
  {Research}},\ }\href@noop {} {\emph {\bibinfo {title} {{Mathematica}}}},\
  \bibinfo {edition} {verion 9.0}\ ed.\ (\bibinfo {address} {Champaign,
  Illinois},\ \bibinfo {year} {2013})\BibitemShut {NoStop}%
\bibitem [{Note1()}]{Note1}%
  \BibitemOpen
  \bibinfo {note} {Notice that if the two fist bins in figure~2 were not
  corrected then the estimated population of $N$ would contain negative
  values.}\BibitemShut {Stop}%
\bibitem [{\citenamefont {Koster}\ \emph {et~al.}(2006)\citenamefont {Koster},
  \citenamefont {Wiggins},\ and\ \citenamefont {Dekker}}]{Koster07022006}%
  \BibitemOpen
  \bibfield  {author} {\bibinfo {author} {\bibfnamefont {D.~A.}\ \bibnamefont
  {Koster}}, \bibinfo {author} {\bibfnamefont {C.~H.}\ \bibnamefont {Wiggins}},
  \ and\ \bibinfo {author} {\bibfnamefont {N.~H.}\ \bibnamefont {Dekker}},\
  }\href {\doibase 10.1073/pnas.0510509103} {\bibfield  {journal} {\bibinfo
  {journal} {Proceedings of the National Academy of Sciences of the United
  States of America}\ }\textbf {\bibinfo {volume} {103}},\ \bibinfo {pages}
  {1750} (\bibinfo {year} {2006})}\BibitemShut {NoStop}%
\bibitem [{\citenamefont {Cao}\ and\ \citenamefont {Li}(2011)}]{Cao2011}%
  \BibitemOpen
  \bibfield  {author} {\bibinfo {author} {\bibfnamefont {Y.}~\bibnamefont
  {Cao}}\ and\ \bibinfo {author} {\bibfnamefont {H.}~\bibnamefont {Li}},\
  }\href {\doibase 10.1021/la104130n} {\bibfield  {journal} {\bibinfo
  {journal} {Langmuir}\ }\textbf {\bibinfo {volume} {27}},\ \bibinfo {pages}
  {1440} (\bibinfo {year} {2011})}\BibitemShut {NoStop}%
\end{thebibliography}

%

\end{document}